\documentclass[lettersize,journal]{IEEEtran}
\pdfoutput=1
\usepackage{amsmath,amsfonts}
\usepackage{algorithmic}
\usepackage{algorithm}
\usepackage{array}
\usepackage{textcomp}
\usepackage{stfloats}
\usepackage{url}
\usepackage{verbatim}
\usepackage{graphicx}
\usepackage{cite}
\usepackage{enumitem}
\usepackage{multirow}
\usepackage{float}
\usepackage{subfigure}
\usepackage{subcaption}
\usepackage{booktabs}
\usepackage{hyperref}
\usepackage{cleveref}
\usepackage{orcidlink}
\begin{document}

\title{Amplify Graph Learning for Recommendation \\ via Sparsity Completion}

\author{Peng Yuan, Haojie Li, Minying Fang, Xu Yu, Yongjing Hao, and Junwei Du
\thanks{}
}

\markboth{}%
{Shell \MakeLowercase{\textit{et al.}}: Amplify Graph Learning framework based on Sparsity Completion}

\IEEEpubid{0000--0000/00\$00.00~\copyright~2024 IEEE}

\maketitle

\begin{abstract}

Graph learning models have been widely deployed in collaborative filtering (CF) based recommendation systems. Due to the issue of data sparsity, the graph structure of the original input lacks potential positive preference edges, which significantly reduces the performance of recommendations. In this paper, we study how to enhance the graph structure for CF more effectively, thereby optimizing the representation of graph nodes. Previous works introduced matrix completion techniques into CF, proposing the use of either stochastic completion methods or superficial structure completion to address this issue. However, most of these approaches employ random numerical filling that lack control over noise perturbations and limit the in-depth exploration of higher-order interaction features of nodes, resulting in biased graph representations. 
  
In this paper, we propose an Amplify Graph Learning framework based on Sparsity Completion (called AGL-SC). First, we utilize graph neural network to mine direct interaction features between user and item nodes, which are used as the inputs of the encoder. Second, we design a factorization-based method to mine higher-order interaction features. These features serve as perturbation factors in the latent space of the hidden layer to facilitate generative enhancement. Finally, by employing the variational inference, the above multi-order features are integrated to implement the completion and enhancement of missing graph structures. We conducted benchmark and strategy experiments on four real-world datasets related to recommendation tasks. The experimental results demonstrate that AGL-SC significantly outperforms the state-of-the-art methods. The code of AGL-SC is available at \url {https://github.com/yp8976/AGL_SC}.
\end{abstract}


\section{Introduction}
\IEEEPARstart{A}{s} an effective solution to the issue of information overload, recommendation systems have been widely implemented in various domains by predicting personalized user preferences. However, in these systems, the predominant issue affecting the accuracy of suggestions stems from the inherent sparsity of user interactions, as it is impractical for users to rate every item. Consequently, researchers have proposed various methods, including content-based and CF-based recommendations \cite{wu2022survey}, aimed at enhancing the model's proficiency in handling sparse data.
\begin{figure*}
    \centering
    \includegraphics[width=0.8\linewidth]{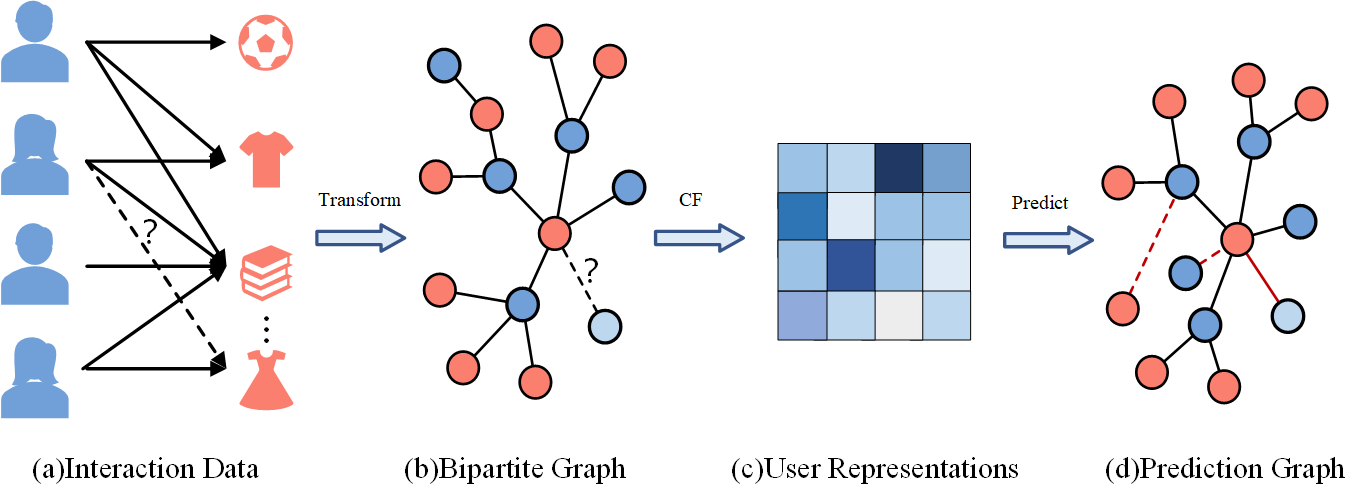}
    \caption{An illustration of the matrix completion algorithm in recommendation systems, demonstrating the transformation from interactions to bipartite graphs for prediction purposes.}
    \label{fig 1}
\end{figure*}

CF-based recommendations predict personalized user preferences by assuming that users with similar behaviors exhibit similar preferences towards items. This approach relies on the observed user-item interactions to learn user and item embeddings. A common paradigm is to parameterize users and items for reconstructing historical interactions, and then predict user preference based on the pairwise similarity between user and item embeddings. Since interactions between users and items can naturally transformed as a user-item bipartite graph, drawing on the success of Graph Neural Networks (GNNs), graph-based CF models have been extensively studied. These models often generate user and item embeddings through recursive message passing operations on the graph structure, which integrate multi-hop neighbors into node representation learning and achieve state-of-the-art recommendation performance. Fig.\ref{fig 1} shows the existing graph collaborative filtering methods, most of graph-based CF models treat all unobserved actions as negative feedback when transforming bipartite graphs. When learning user representations, these models often study propagate interactions among neighbors by stacking GNN layers, utilizing the graph structure and message passing mechanisms. However, this approach causes the lack of potential positive preference links and overlooks the relevance of high-order neighbors among users and items. Coupled with the excessively sparse interaction data and substantial noise in practice, these issues become more pronounced. Such conditions lead to significant biases in graph node representations \cite{ye2023grace}, ultimately leading to poor performance.

\IEEEpubidadjcol

Matrix completion has been widely applied in recommendations based on CF as a technique used to reconstruct lost data in sparse matrices \cite{xie2021deep,chen2022review}. Early matrix completion technologies, such as Singular Value Thresholding (SVT), treated user-item implicit feedback as a low-dimensional user-item similarity matrix for learning. Due to the limitations of linear transformations, these early approaches faced relatively high computational costs when dealing with large datasets \cite{farias2022uncertainty}. With the development of deep learning, advanced recommendation models have adopted the matrix completion concept, further employing graph neural networks for collaborative filtering to enhance recommendation performance. Such models transform recommendation tasks into issues of graph node representation and edge prediction, by propagating information in the form of graph between users and items \cite{he2024kgcna}. For instance, IGMC proposed an inductive matrix completion approach, achieving edge prediction based on subgraph extraction methods \cite{Zhang2020Inductive}. EGLN integrates the concept of mutual information maximization, iteratively propagates information between nodes in graph to realize edge prediction \cite{yang2021enhanced}. These methods mitigate data sparsity and enhance the graph representation, thus improving the quality of recommendations. However, the incorporation of noise is random, which may lead to poor local-global consistency in the graph structure, and could potentially lead to implicit changes in the model's representations in latent space, such as posterior collapse \cite{dai2020usual}.

Despite matrix completion techniques have been effective in dealing with the sparse data in recommendation systems, graph-based CF models still grapple with the issue of bias in long-tail distributions. Inspired by the success of generative self-supervised learning in CV, the generative models can well reconstruct the input graph without information distortion. Among them, VAE was proposed as a perturbation data augmentation method introduced into the graph representation learning for recommendation systems. It has significant potential in modeling user and item features, as well as debiasing \cite{he2022masked,wu2020diffnet++,nakagawa2022gromov}. In this paper, we leverage variational inference to reconstruct the graph structure without information bias.

To this end, in this paper, we propose the AGL-SC strategy based on sparse completion for enhanced graph learning. Addressing the issues of representation bias caused by data sparsity and noise, we generate feature representations of user and item nodes through graph neural networks. excavating direct interactions of user and item nodes as superficial structural features. Simultaneously, by focusing on similar user and item groups, we employ an improved factorization method to mine high-order interaction features, thereby enhancing the model's ability to distinguish similar user representations \cite{li2023generalized}. Finally, these representations are fed into a variational generative function for multi-order feature collaborative inference, completing the complex nonlinear interaction features of users. This approach results in more precise and comprehensive user recommendations, and offers strategies for mitigating representation bias in various model nodes.

The primary contributions are as follows:

\begin{itemize}[leftmargin=*]
  \item We propose a sparse completion framework based on VAE that captures the latent multi-order interaction distribution features of user-item pairs. This framework generates comprehensive distributions of user preferences, thereby enhancing graph representation.
  \item We present an update mechanism that integrates high-order interaction features of user-item pairs. This mechanism effectively merges the characteristics of high-order interactions between users and items with generated latent vectors.
  \item Extensive experiments conducted on four datasets, compared with other baseline models. The proposed AGL-SC model demonstrated significant performance improvements.
\end{itemize}

\section{Related Work}

\subsection{Graph Learning for Recommendation}

To enhance the efficiency of recommendation systems, existing research has inclined towards using graph learning to process unstructured graph data. The primary objective is to efficiently transform graph data into low-dimensional, dense vectors, ensuring that the graph's informational and structural properties are precisely mapped in the vector space.

Initially, graph learning based on node embedding through matrix decomposition techniques like Neural Matrix Factorization (NMF) \cite{he2017neural2} and Singular Value Decomposition (SVD), converting nodes into a low-dimensional vector space. However, these methods were characterized by high time and space complexity. Subsequently, with the success of word vector methodologies in Natural Language Processing (NLP), some approaches, such as DeepWalk \cite{perozzi2014deepwalk} and Node2Vec \cite{grover2016node2vec}, facilitated large-scale graph learning by converting graphs into sequences. Yet, this method also led to a suboptimal utilization of the inherent structural information of the graphs.

In recent years, drawing on the principles of CNN, RNN, and autoencoders, researchers have developed Graph Neural Network architectures,which have demonstrated outstanding performance in structural information extraction and node representation. Contemporary research has largely focused on space-based Graph Neural Networks for recommendations \cite{shi2021sgcn,ren2023disentangled,yang2023generative}. PinSage \cite{ying2018graph}, a content recommendation model based on Graph Neural Networks, propagates item features through an item-item association graph. NGCF \cite{wang2019neural} employs Graph Neural Networks to model high-order collaborative signals between users and items during the embedding learning process. LightGCN \cite{he2020lightgcn} introduces neighborhood aggregation, propagating embeddings across user-item interaction graphs and simplifies the model by removing feature transformation and non-linear activation. UltraGCN \cite{mao2021ultragcn} adjusts the relative importance of different types of relationships flexibly using edge weight allocation in a constraint loss. SVD-GCN \cite{peng2022svd}, replacing part of the Graph Neural Network methods with SVD, utilizes the top k singular vectors for recommendations. These methods optimize information transmission and node representation through various strategies, thereby improving recommendation efficiency and precision. However, the information transmission mechanism of Graph Neural Networks may slow down network convergence, inherit and amplify noise within the network \cite{ye2023grace}, leading to bias in the embedding of graph structure nodes and over-smoothing issues.

\subsection{Matrix Completion}

Matrix completion, based on the learning of graph-structured data, interprets users' ratings or purchasing behaviors towards items as links in a graph. By filling in the missing edges in the interaction graph, it facilitates the recommendation tasks \cite{chen2020efficient,cao2021bipartite}. In recommendation systems, users often express their preferences indirectly through implicit feedback, such as browsing product interfaces, rather than through direct actions like ratings or likes. These behaviors, serving as implicit feedback in recommendation systems, can more accurately reflect user preferences. Such actions are crucial for modeling user preferences and enhancing the effectiveness of recommendations \cite{xie2021deep}. Therefore, how to complete user preferences from implicit feedback has become an important academic topic.

Recent research tends to incorporate user-item interaction features as auxiliary information for completion. Monti et al. introduced user and item neighborhood networks MGCNN to extract latent features of users and items \cite{monti2017geometric}, while Berg et al. utilized one-hot encoding of node IDs for feature extraction by GC-MC \cite{berg2017graph}. Nguyen et al. implemented a deep model-based Conditional Random Field algorithm DCMC to handle large-scale \cite{nguyen2019geometric}, highly incomplete data. Wu et al. proposed a completion model AGCN \cite{wu2020joint}, that integrates user-item attribute information. Zhang et al. introduced MC2G \cite{zhang2022mc2g}, which combines social networks and item similarity graphs to enhance recognition of user preferences.

However, each of these methods has its limitations. MGCNN relies on the graph Laplacian operator, limiting its generalization capabilities on new graphs. GC-MC and DCMC use linear decoders or simple neural networks for completion prediction, might oversimplify actual user behavior patterns. These models are based on optimizing discriminative strategies with given user interaction information that overly simplify user behavior. Introducing additional signals, such as attribute description text, can increase model complexity and computational costs, potentially introducing more noise and irrelevant information \cite{wang2023diffusion}. Moreover, these models, typically based on fixed-length vectors, lack the capacity to represent the diversity and uncertainty of user interests \cite{yang2023generate}.

Our work employs generative models, which infer the interaction probabilities for items that users have not interacted with by sampling from learned probability distributions \cite{ye2023graph}. In practical applications, the smoothness of VAEs in the distribution helps to avoid drastic fluctuations in model parameters due to minor differences in samples, even when user behavior undergoes subtle changes \cite{liang2018variational}. This capability enables us to better reconstruct missing node features, explore, and complete user preferences.

\section{THE PROPOSED MODEL}

\begin{figure*}
    \centering
    \hspace{17pt} 
    \includegraphics[width=0.85\linewidth]{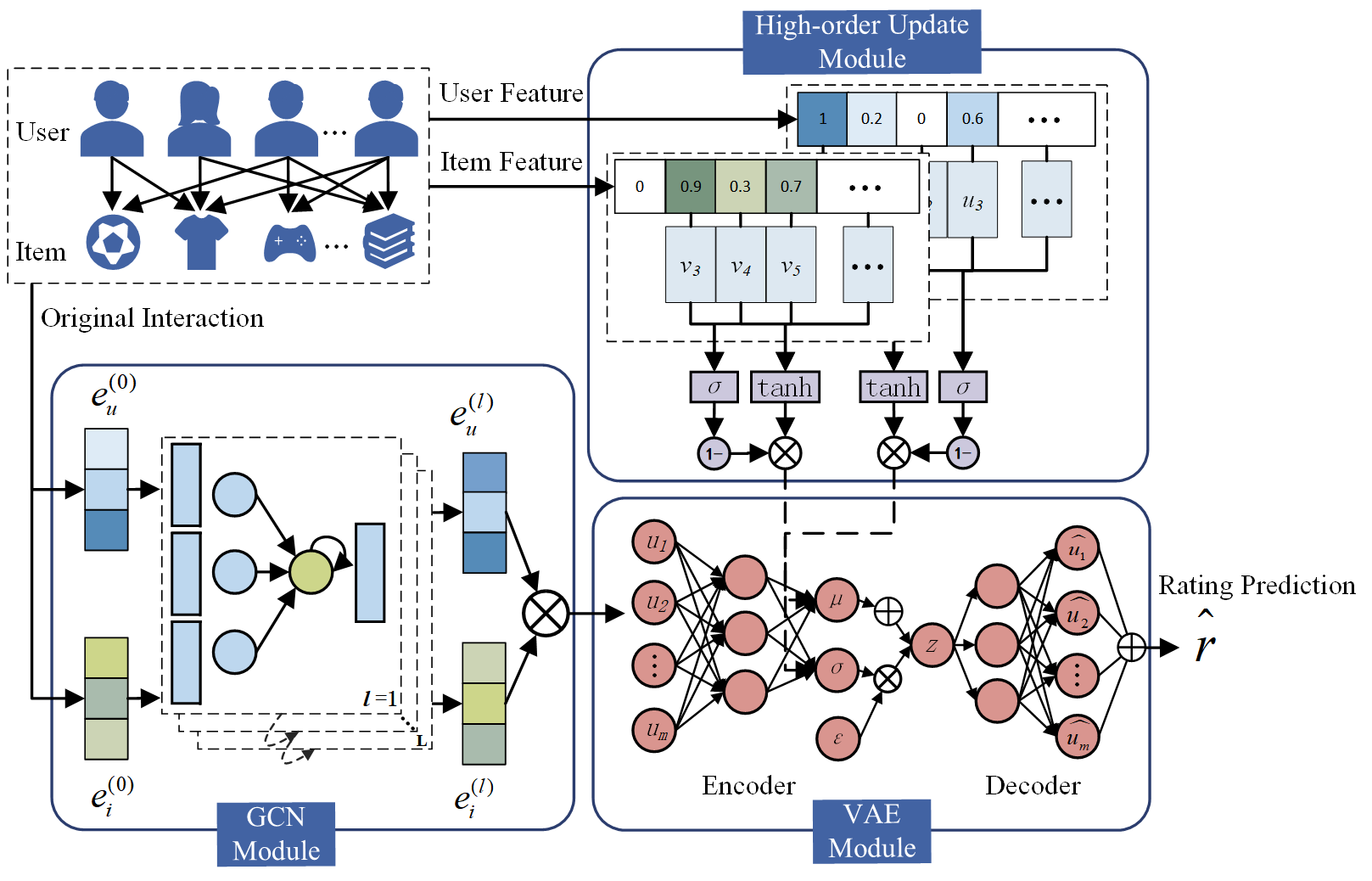}
    \caption{The illustration of our proposed AGL-SC framework, which
consists of a GCN module for initial feature extraction from original user-item interactions, high-order update module for incorporating high-order constraints, and a VAE module for generating predictions.}
    \label{fig 2}
\end{figure*}

We proposed Amplify Graph Learning framework based on Sparsity Completion (AGL-SC), which comprises three main components: 1) The graph learning module maps user-item bipartite graphs into low-dimensional feature vectors for both user and item nodes; 2) The high-order constraint module derives high-order interaction feature vectors for nodes through factorization, transforming them into perturbation noise for VAE, thereby facilitating improved graph structure completion and node embedding learning; 3) In the VAE module, a multinomial likelihood function is incorporated to utilize multi-order features for variational inference and prediction of the probabilistic distribution of user preferences, aiming to generate a more comprehensive understanding of user preferences and enhance recommendation performance. For a better illustration, we show the overall framework of AGL-SC in Fig.\ref{fig 2}.

\subsection{User-Item Feature Learning} \label{Feature Learning}

The graph learning module accomplishes the embedding of node features within the graph by modeling the direct interaction structures between user and item nodes.

In recommendation systems, there are two primary entities: a set of users ${U}(|{U}|={M})$ and a set of items ${V}({V}|={N})$. We represent the implicit feedback of user-item interactions using an interaction matrix ${{R}}\in{{{\mathbb{R}}}}^{M\times N}$ where if user ${u}$ interacts with item ${i}$, the corresponding element is $\mathbf{r}_{ui}=1$ otherwise $\mathbf{r}_{ui}=0$. Most neural graph recommendation models define the user-item graph as $\mathcal{G}=\{{U}\cup{V},{A}\}$, wherein the adjacency matrix ${A}$ is an unweighted matrix, defined as follows:
\begin{equation}
  A=\left[
    \begin{array}{llllllllll}
      0     & R \\
      {R}^T & 0
    \end{array}\right].
\end{equation}

Utilizing initialization, all vertices in the network are mapped into a low-dimensional embedding space $\mathbf e_u\in \mathbb{R}^{M\times d}$ and $\mathbf e_v\in \mathbb{R}^{N\times d}$, transforming them into vector matrices to obtain the initial embedding representations of users and items. $D$ is a diagonal node degree matrix, ${E}\in\mathbb{R}^{(M+N)\times d}$ represents the embedding matrix. Subsequently, these initialized embeddings of users and items are fed into a graph neural network to learn model parameters. By stacking ${l}$ layers of embedding propagation, users (and items) can receive messages ${H}$ propagated from their ${l}$ order neighbors:
\begin{equation}
  \begin{array}{l}
    H^l=\widetilde{A}^lE ,\\
    \widetilde{A}=D^{-\frac{1}{2}}AD^{-\frac{1}{2}}.
  \end{array}
\end{equation}

The feature representations of users and items are calculated comprehensively, and the node embeddings of both at each layer are accumulated and aggregated through a pooling function. Subsequently, an inner product operation is utilized to derive the latent rating matrix ${P}$ :
\begin{equation}
  \mathbf{e}=p o o l i n g_{AVG}(H^{l}|l=0,1,...,L).
\end{equation}
\begin{equation}
  {P}=\mathbf{e}_{u}^{T} \cdot \mathbf{e}_{v}.
\end{equation}

\subsection{Implicit High-Order Feature Constraint} \label{High-Order Feature}

Accurately mining the structural constraints of a bipartite graph is a pivotal approach to enhancing bipartite graph learning \cite{wang2023collaboration,jiang2019semi}. Homogeneous node high-order constraints are employed to capture the inherent similarities among users or items, while heterogeneous node interaction high-order constraints are utilized to gauge the extent of a user's preference for an item. The high-order constraint module, drawing upon factor decomposition methods, extracts high-order features of nodes and amalgamates both types of node constraints to enhance the accuracy of the model's node representation. Building on this, we initially derive the user feature matrix ${E}_{U}\in \mathbb{R}^{M\times d}$ and the item feature matrix ${E}_{V}\in \mathbb{R}^{N\times d}$ by transforming the embeddings into dense vector representations. Then for each user-item instance $(i,j)$, a mapping function is applied as:
\begin{equation}
  \phi(E_{ui}, E_{vj}) = E_{ui} \odot E_{vj},
\end{equation}

where $\odot$ denotes the element-wise product of vectors. We then project the vector to the next layer:
\begin{equation}
  \widehat{R}_{ij} = sigmoid \left( h^T \left(E_{ui} \odot E_{vj} \right) \right),
\end{equation}

where $\mathbf{h}$ denotes edge weights of the output layer. The calculated interaction strengths of users to various items are utilized as weights to derive an aggregated feature matrix that expresses the similarity of interactions among users. Similarly, an aggregated feature vector matrix is constructed to represent the similarity between items:
\begin{equation}
\begin{aligned}
  \widehat{{{W}}}_{U} = E_ {U}(\widehat RE_{V})^T, \\
  \widehat{{{W}}}_{V} = E_ {V}(\widehat RE_{U})^T.
\end{aligned}
\end{equation}

Furthermore, the similarity feature vector matrix is utilized to identify groups of users and items with similar preferences. we derives the relationship matrices between users ${W}_{U}$ and between items ${W}_{V}$ through the interaction matrix ${R}$:
\begin{equation}
  {W}_{U}=R{R}^{T} , {W}_{V}={R}^{T}R.
\end{equation}

By minimizing the Kullback-Leibler Divergence between these distributions, the embedding vector representations of the nodes are optimized. Consequently, the high-order constraint loss function for the nodes is defined as follows:
\begin{equation}
  {\mathcal{L}} _ {MC} = {D_{KL}}({W_{U}}||{\widehat{W}_{U}}) + {D_{KL}}({W_{V}}||{\widehat{W}_{V}}).
\end{equation}

\begin{equation}
  \footnotesize
  \mathcal{L}_{{MC}} = \sum_{i,j=1}^{M} W_{Uij} \left( \log \frac{W_{Uij}}{\widehat{W}_{Uij}} \right) 
  + \sum_{i,j\in {V}}^{N} W_{Vij} \left( \log \frac{W_{Vij}}{\widehat{W}_{Vij}} \right).
\end{equation}

\subsection{Structure Generative Learning}

In recommendation systems, VAEs predominantly utilize user distributions for generation, adeptly capturing the latent relationships between users and items \cite{ma2019learning,shenbin2020recvae}, especially in handling sparse data. We design a generative module to enhance the completion of missing graph structures, thereby strengthening the model's understanding and prediction of complex user behavior patterns.

VAEs exhibit remarkable performance in handling sparse data, a particularly pertinent feature for addressing the common issue of data sparsity in recommendation tasks. In recommendation systems, VAEs predominantly generate recommendations based on user distributions, effectively capturing the latent relationships between users and items, thereby providing more precise and personalized recommendations. Numerous studies input the missing graph structures directly into VAEs to tackle recommendation issues under data sparsity conditions. However, this approach lead to an increase in the neural network's capacity, making it challenging for the VAE to capture detailed feature information. Consequently, this generative module can adjusting the intensity of noise to enhance the missing graph structures, thereby improving the relevance and diversity of recommendations.

In Section \ref{Feature Learning}, we have already fitted the feature representations of nodes using graph neural networks and obtained the predicted matrix $\widehat R$. In order to better construct the distribution of latent variables in the generative module, the latent matrix $P$ of nodes are optimized using Wasserstein distance \cite{tolstikhin2017wasserstein}:
\begin{equation}
  {\mathcal{W}}[\widehat R,P]=\underset{{\gamma\in\Gamma(X\sim \widehat R,Y\sim P)}}{\mathrm{\inf}}\mathbb{E}_{{(X,Y){\sim}\gamma}}\mathsf{[||}X-Y{||]},
\end{equation}

where $X$ and $Y$ represent random variables drawn from two probability distributions $\widehat R$ and $P$, respectively. VAE perceives the latent vector as a probabilistic distribution, with the output of its hidden layer encompassing two dimensions,  $\mu$ and $\sigma$, which represent the mean and variance of the normal distribution parameters that the encoding $\mathbf{z}$ adheres to. This facilitates a smooth transition of vectors among similar samples. We proceed with computations utilizing the optimized distribution ${P}$ for user $u$:
\begin{equation}
\begin{array}{l}
  {q_\phi }(\mathbf z|P) = \prod\nolimits_{u = 0}^M {{q_\phi }({\mathbf z_u}|{P_u})} \;,\\
  {q_\phi }({\mathbf z_u}|{P_u}) = N({\mathbf z_u};{\mu _\phi }(u),diag(\sigma _\phi ^2(u)),

\end{array}
\end{equation}
where $\mathbf z_{u}$ and ${P}_{u}$ respectively represent the user's low dimensional structural embedding and interaction prediction. $\mu$ and $\sigma$ denote the mean and variance of the latent Gaussian distribution generated by the encoder, with $\phi$ symbolizing the training parameters of the encoder, estimated by a parametric encoder network. To ensure that the distribution closely approximates the true posterior distribution, a parametric encoder network is employed to minimize the Kullback-Leibler Divergence between the two distributions:
\begin{equation}
  \phi^{*}=\arg\min_{\phi}\operatorname{D_{KL}}(q_{\phi}(\mathbf z_{u}\mid{P}_{u}),p_{\theta}\left(\mathbf z_{u}\right)).
\end{equation}

Subsequently, we have devised a method through gated units to filter and integrate input features, thereby controlling the impact of the latent high-order constraints computed in Section \ref{High-Order Feature} on the generation of the graph structure. We denote by $\mathbf{x}_{u}$ the latent vector of user ${u}$, ${h}$ refers to the representation of the hidden state. The gating signal for each user can be expressed as follows:
\begin{equation}
  \begin{array}{l}
    {{\mathbf h_{\mu_u}=\,tanh(W_{\mu_u}\cdot \mathbf x_{\mu_u})}},          \\
    {{\mathbf h_{\sigma_u}=\,tanh(W_{\sigma_u}\cdot \mathbf x_{\sigma_u})}}, \\
    \mathbf z=sigmoid(W_z\cdot[\mathbf x_{\mu_u},\mathbf x_{\sigma_u}]),
  \end{array}
\end{equation}

where ${W} \cdot$ is a learnable transformation matrix8. Calculate the enhanced distribution representation for each user ${u}$ by the following method:
\normalsize
\begin{equation}
  \begin{array}{l}
    \mu_{u}=\mathbf z\cdot \mathbf h_{\mu_{u}},           \\
    \sigma_{u}=(1-\mathbf z)\cdot \mathbf h_{\sigma_{u}}, \\
    p({\widehat{P}}_{u})={\mathcal{N}}\left(\mu_{u},d i a g(\sigma_{u}^{2})\right).
  \end{array}
\end{equation}

The decoder reconstructs the interaction matrix through conditional probability. Consequently, the Kullback-Leibler Divergence between this and the posterior distribution can be calculated by the following formula:
\begin{equation}
  \scriptsize
  {D_{KL}}({q_\phi }({\mathbf z_u}|{\widehat P_u})||{p_\theta }({\mathbf z_u})) = \int {{q_\phi }({\mathbf z_u})(\log {p_\theta }({\mathbf z_u}) - \log {q_\phi }({\mathbf z_u})){{d}}z}.
\end{equation}

Finally, to obtain targeted enhanced feature representations, sampling from the posterior distribution yields a new latent layer encoding $\mathbf z$, which is then inputted into the generative network. This approach allows us to preserve the inherent structure in the process of generating the graph structure, while also enhancing the model's generalization capabilities:
\begin{equation}
  \scriptsize
  {{\mathcal L}_{VAE}} = {\mathbb{E}_{{q_\phi }({\mathbf z_u}|{{\widehat P}_u})}}\left[ {\log {p_\theta }({\mathbf z_u}|{{\widehat P}_u})} \right] - {D_{KL}}({q_\phi }({\mathbf z_u}|{\widehat P_u})||{p_\theta }({\mathbf z_u})).
\end{equation}

\subsection{Loss Function}

In this paper, we employ the Bayesian Personalized Ranking (BPR) loss for recommendations, aiming to optimize recommendations by maximizing the predicted ranking of items with which a user has already interacted (positive samples) compared to those they have not (negative samples). Specifically, this objective is achieved by optimizing the following loss function:
\normalsize
\begin{equation}
  {\mathcal{L}}_{REC}=\ \sum_{u\in U,i,j\in V}\,-\,\ln\,sigmoid({\widehat{{P}}}_{u i}-{\widehat{{P}}}_{u j}).
\end{equation}

Under the influence of the VAE, the model's capacity to extract implicit feedback from users is adjusted through the decoder. An additional loss term is introduced to control the gated units' ability to capture the high-order connectivity of user preferences, thereby modeling the complex relationships between users and items more precisely and enhancing the performance of the model's recommendations.

We fine-tune the parameters to balance recommendation loss and generative loss for optimal results. The final loss function can be expressed as follows:
\begin{equation}
  {{\mathcal L}_{AGL-SC}} = {{\mathcal L}_{REC}}+{\lambda}{{\mathcal L}_{MC}}+{\beta}{{\mathcal L}_{VAE}}.
\end{equation}

\begin{table*}[ht]
  \centering
  \caption{Overall performance comparison. Improv. denotes the relative improvements over the best baselines.}
  \label{table 1}
  \begin{tabular}{lcccccccccccc}
    \midrule
    \multirow{2}{*}{Model} & \multicolumn{3}{c}{Amazon-Electronics} & \multicolumn{3}{c}{ML-100K} & \multicolumn{3}{c}{ML-1M} & \multicolumn{3}{c}{Yelp}                                                                                  \\
    \cmidrule{2-4} \cmidrule{5-7} \cmidrule{8-10} \cmidrule{11-13}
    & R@20 & N@20 & N@40   & R@20   & N@20   & N@40   & R@20   & N@20   & N@40  & R@20 & N@20 & N@40\\
    \hline
    DeepWalk  &0.0725 &0.0579 &0.0654 & 0.2832 & 0.2371 & 0.2758 & 0.1348 & 0.1057 & 0.1626 & 0.0476 & 0.0378 & 0.0452 \\
    Node2Vec  &0.0799 &0.0534 &0.0724 & 0.3005 & 0.2568 & 0.2974 & 0.1475 & 0.1186 & 0.1677 & 0.0452 & 0.0360 & 0.0552 \\
    \midrule
    NGCF      &0.1286 &0.0710 &0.0830 & 0.3413 & 0.3034 & 0.3341 & 0.2475 & 0.2457 & 0.3028 & 0.0579 & 0.0477 & 0.0602 \\
    LightGCN  &0.1259 &0.0715 &0.0855 & 0.3422 & 0.3097 & 0.3348 & 0.2511 & 0.2507 & 0.3072 & 0.0617 & 0.0503 & 0.0631 \\
    DGCF      &0.1298 &0.0714 &0.0853 & 0.3433 & 0.3093 & 0.3386 & 0.2576 & 0.2504 & 0.3069 & 0.0649 & 0.0530 & 0.0695 \\
    UltraGCN  &0.1291 &0.0791 &0.0894 & 0.3445 & 0.3124 & 0.3574 & 0.2640 & 0.2513 & 0.3092 & 0.0654 & 0.0534 & 0.0682 \\
    SVD-GCN   &0.1318 &0.0809 &0.0913 & 0.3505 & 0.3145 & 0.3601 & 0.2737 & 0.2552 & 0.3112 & 0.0683 & 0.0561 & 0.0719 \\
    VGCL      &0.1362 &0.0824 &0.0935 & 0.3524 & 0.3152 & 0.3606 & 0.2756 & 0.2598 & 0.3120 & 0.0687 & 0.0565 & 0.0707 \\

    \midrule
    \bfseries{AGL-SC(Pytorch)}                 
    & \bfseries{0.1413} & \bfseries{0.0850} & \bfseries{0.0967} & \bfseries{0.3598} & \bfseries{0.3226} & \bfseries{0.3683} 
    & \bfseries{0.2834} & \bfseries{0.2653} & \bfseries{0.3164} & \bfseries{0.0703} & \bfseries{0.0587} & \bfseries{0.0742} \\

    \bfseries{AGL-SC(MindSpore)}                 
    & \bfseries{0.1139} & \bfseries{0.0459} & \bfseries{0.0671} & \bfseries{0.3506} & \bfseries{0.3069} & \bfseries{0.3233} 
    & \bfseries{0.2551} & \bfseries{0.2671} & \bfseries{0.2787} & \bfseries{0.0701} & \bfseries{0.0479} & \bfseries{0.0655} \\

    \midrule
    \bfseries{{\%}Improv.}
    & 3.74{\%} & 3.16{\%} & 3.42{\%} & 2.10{\%} & 2.35{\%} & 2.14{\%} & 2.83{\%} & 2.12{\%} & 1.4{\%} & 2.34{\%} & 3.89{\%} & 3.20{\%}\\
    
    \bfseries{${p}$-value} 
    & 2.81e-12 & 4.59e-9 & 3.84e-9 & 1.71e-9 & 3.72e-8 & 1.24e-9 & 1.02e-6 & 6.84e-6 & 9.02e-8 & 9.64e-9 & 1.25e-10 & 2.08e-7\\
    \hline
  \end{tabular}
\end{table*} 

\section{EXPERIMENTS}

To demonstrate the effectiveness of AGL-SC, we have conducted extensive experiments to address the following research questions:
\begin{itemize}[leftmargin=*]
  \item {\bfseries RQ1}: Does our proposed method enhance recommendation accuracy more effectively than other baseline methods?
  \item {\bfseries RQ2}: What impact do the various modules of the model have on its final performance?
  \item {\bfseries RQ3}: How do various hyperparameter settings affect the model's performance?
  \item {\bfseries RQ4}: Does our method effectively improve the model's extraction of implicit feedback, thereby enhancing recommendation performance?
\end{itemize}

\subsection{Experimental Settings}

\subsubsection{{\bfseries Datasets}}

To evaluate the effectiveness of our model, we use four publicly available datasets of different scales, including MovieLens100K,
MovieLens1M, Amazon Electronics, and Yelp. Table \ref{table 2} presents detailed information corresponding to these datasets.
\begin{table}
\centering
  \caption{Statistics of the experimented datasets.}
  \label{table 2}
  \begin{tabular}{@{}ccccc@{}}
    \toprule
    Dataset           & \#Users & \#Items & \#Interactions & Density \\
    \hline
    Amazon-Electronics & 1435    & 1522    & 35,931         & 1.645\% \\
    MovieLens100K     & 943     & 1674    & 55,375         & 3.507\% \\
    MovieLens1M       & 6022    & 3043    & 995,154        & 5.431\% \\
    Yelp              & 31668   & 38048   & 1,561,406      & 0.130\% \\
    \bottomrule
  \end{tabular}
\end{table}

\subsubsection{{\bfseries Evaluation Metric}}

To evaluate the performance of various methods, we employed the rank-based metric Normalized Discounted Cumulative Gain NDCG@K and the relevance-based measure Recall@K.

NDCG@K:
\begin{equation}
  \begin{array}{l}
    {{DCG}} = \mathop \sum \limits_{i = 1}^K \frac{{re{l_i}}}{{{{\log }_2}(i + 1)}},  \\
    IDCG = \mathop \sum \limits_{i = 1}^K \frac{{re{l_i}^ * }}{{{{\log }_2}(i + 1)}}, \\
  \end{array}
\end{equation}
where ${rel}_{i}$ represents the relevance of item ${i}$ for the target user, and is the relevance of the item ideally ranked at position ${i}$. DCG and IDCG respectively denote the computed Discounted Cumulative Gain and the ideal Discounted Cumulative Gain.

Recall@K:
\begin{equation}
  {\mathop{Re}\nolimits} call = \frac{{TP}}{{TP + FN}},
\end{equation}
where TP signifies the true positives, accurately predicted as such, where the actual value is positive, and the predicted value is also positive; FN represents the false negatives, incorrectly predicted as such, where the actual value is positive, but the prediction erroneously indicates a negative.

\subsubsection{{\bfseries Baselines}}

To validate the efficacy of our proposed recommendation method, we compared it with several comparable state-of-the-art model approaches. These include graph embedding methods (DeepWalk, Node2Vec), and graph collaborative filtering-based methods (NGCF, LightGCN, DGCF, UltraGCN, SVD-GCN). For each baseline method, we adjusted their parameters, on the same dataset, reported their best-performing parameters for comparison with our model.

\noindent \textbf{Graph embedding-based methods:}
\begin{itemize}[leftmargin=*]
  \item {\bfseries DeepWalk} \cite{perozzi2014deepwalk}: Transforms graph structure data into a natural language processing task using a random walk strategy to generate node embeddings in the graph.
  \item {\bfseries Node2Vec} \cite{grover2016node2vec}: Combines DeepWalk's random walk approach with a custom exploration-exploitation strategy to more flexibly capture homophily and structural equivalence in networks.
\end{itemize}  

\noindent \textbf{Graph collaborative filtering-based methods:}
\begin{itemize}[leftmargin=*]  
  \item {\bfseries NGCF} \cite{wang2019neural}: Leveraging graph neural networks to enhance collaborative filtering, it refines user and item embeddings by incorporating high-order connectivity relations and aggregates neighbor embeddings from different propagation layers to learn representations of users and items.
  \item {\bfseries LightGCN} \cite{he2020lightgcn}: Simplifies the design of GCN to make it more concise and suitable for recommendations, learning user and item embeddings through linear propagation on the user-item interaction graph, and employing a weighted sum of embeddings learned across all layers as the final embedding. We set three layers as the baseline.
  \item {\bfseries DGCF} \cite{ren2023disentangled}: Pays particular attention to the fine granularity of user-item relationships in terms of user intent. Models the intent distribution for each user-item interaction and iteratively refines intent-aware interaction graphs and representations. We set four intents as the baseline.
  \item {\bfseries UltraGCN} \cite{mao2021ultragcn}: An extension of LightGCN, it builds a loss function by setting a convergence mode to avoid problems with multi-layer stacking. We set gamma=1e-4, lambda=1e-3 as the baseline.
  \item {\bfseries SVD-GCN} \cite{peng2022svd}: Replaces the core design in the GCN method with a flexible truncated SVD, utilizing the top K singular vectors for recommendation purposes. We set alpha=3, beta=2 as the baseline.
  \item {\bfseries VGCL} \cite{yang2023generative}: Integrates variational graph generation with contrastive learning, predicting user preferences using a contrastive loss based on variational graph reconstruction. We set t=0.25, lambda=0.1, gamma=0.5 as the baseline.
\end{itemize}

\subsubsection{{\bfseries Parameter Settings}}

We implemented the AGL-SC model and all baselines using PyTorch, employing Kaiming initialization to process user-item embeddings, resulting in 128-dimensional initial embeddings. Within the model, the number of GCN layers is set to 2, and the VAE hidden state vector dimension is 200. During training, a dropout rate of 0.2 is used to mitigate the likelihood of overfitting. The network optimization employs the Adam optimizer, with an initial learning rate of 0.001. The training is set for a maximum of 100 epochs. For evaluation, if there is no improvement over 10 consecutive training epochs, the training is terminated. Moreover, we reproduce the results on MovieLens1M dataset with MindSpore framework (Huawei 2024).

\subsection{Overall Performance Comparisons}

Table \ref{table 1} showcase the overall performance of various baseline models across four datasets. It is observable that our proposed AGL-SC model outperforms the other baseline models in all metrics.We found that methods based on graph neural networks achieved better performance than graph embedding methods, as the two graph embedding models solely use the user-item interaction bipartite graph as input. This highlights the ability of graph learning to capture user preferences by modeling high-order user-item graph structures, compensating for the models' capability to handle implicit feedback and alleviating the issue of over-smoothing. This is particularly evident when dealing with users or items of high similarity, where the model can more accurately identify and emphasize important connections while reducing the impact of noise.

Among the existing graph recommendation methods, VGCL \cite{yang2023generative} achieved the best baseline performance, followed by SVD-GCN \cite{peng2022svd} and UltraGCN \cite{mao2021ultragcn}. VGCL, by utilizing variational graph reconstruction, addresses the limitations of current contrastive learning view enhancement methods, thus improving recommendation performance. This underscores the effectiveness of using generative models in recommendation systems to yield richer and more diverse recommendations, with variational graph reconstruction offering a better graph structure than simple data enhancement. SVD-GCN and UltraGCN, building upon the LightGCN \cite{he2020lightgcn} framework, proposed simplification strategies for graph learning to denoise user interactions. UltraGCN introduced item similarity into graph recommendations, while SVD-GCN added both item and user similarity loss. These comparisons demonstrate that incorporating high-order feature constraints in graph recommendations can effectively enhance recommendation performance.

Beyond the aforementioned factors, our model through a combination of graph neural networks and a sparse completion strategy of generative models. In processing sparse data, the generative module effectively infers missing interactions, enhancing the predictive accuracy of the model. Additionally, the high-order constraint module strengthens the filtration of irrelevant data noise, ensuring the relevance of recommendations, thereby achieving optimal performance. Furthermore, when comparing our model with graph neural network-based models like LightGCN and UltraGCN, we observed that AGL-SC improved metrics across all four datasets, although the specific improvements varied per dataset. We believe these differences reflect the model's effectiveness in completing missing graph structures, a capability not effectively addressed by other baseline models in situations of missing interactions.
\subsection{Ablation Study}

\begin{figure}[htb]
  \centering
  \includegraphics[width=0.43\textwidth]{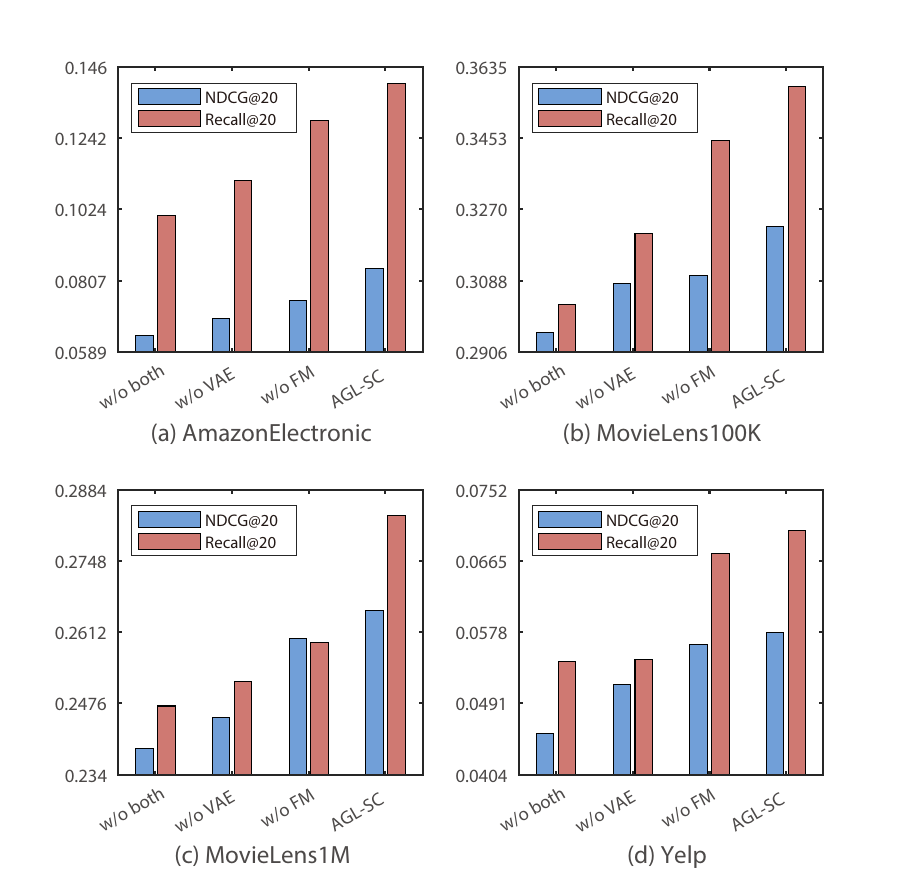}
  \caption{Performance of ablated models on datasets in terms of Recall@20 and NDCG@20.}
  \label{fig 3} 
\end{figure}

We introduced multiple components to enhance the model's performance in recommendation tasks. To assess the importance and contribution of each component, we sequentially removed components and compared the performance of AGL-SC and its corresponding variants in terms of the recommendation performance metrics Recall@20 and NDCG@20, as shown in Table \ref{table 3} and Fig.\ref{fig 3} for a better illustration.

  {\bfseries w/o VAE}: This variant removes the generative module of the model, using a combination of graph neural networks and factorization machines to train embeddings, in order to evaluate the role of the generative module in model performance.

  {\bfseries w/o FM}: This variant removes the model's high-order constraint module, relying solely on embeddings trained using graph neural networks, to explore the effect of introducing high-order features of user-item interactions on enhancing prediction accuracy.

  {\bfseries w/o both}: This is the model's basic version, retaining only the GNN module of our model.

\begin{table}
\centering
  \caption{Performance comparison among DGCF and different AGL-SC variants.}
  \label{table 3}
  \begin{tabular}{ccccc}
    \toprule
    \multirow{2}{*}{Model} & \multicolumn{2}{c}{MovieLens1M} & \multicolumn{2}{c}{Yelp}                   \\
    \cmidrule{2-3} \cmidrule{4-5}
                           & R@20                            & N@20                     & R@20   & N@20   \\
    \hline
    DGCF               & 0.2576                      & 0.2504                   & 0.0649 & 0.0530 \\
    ${ w/o }$ both         & 0.2471                          & 0.2390                   & 0.0542 & 0.0454 \\
    ${ w/o }$ VAE          & 0.2517                          & 0.2449                   & 0.0544 & 0.0514 \\
    ${ w/o }$ FM           & 0.2593                          & 0.2599                   & 0.0674 & 0.0563 \\
    AGL-SC(Pytorch)                 & 0.2834                          & 0.2653                   & 0.0703 & 0.0587 \\
    AGL-SC(MindSpore)                 & 0.2551                          & 0.2671                   & 0.0655 & 0.0479 \\
    \bottomrule
  \end{tabular}
\end{table}

After removing the generative module, the overall performance of the model declined. However, compared to general models like NGCF, the metrics were largely on par or even slightly improved. We attribute this to the high-order constraint module's capability for deep feature mining of bipartite graph data, an aspect often overlooked in standard graph neural network architectures, especially in adequately modeling the complex relationships between user-user and item-item pairs.

After removing the high-order constraint module, the model still exhibited superior performance compared to DGCF. This underscores the efficacy of using VAE generative models for structural enhancement and completion in recommendation systems. Unlike models that merely execute encoding compression, VAEs ensure the integrity of information. This characteristic is particularly crucial for recommendation systems, as it guarantees that the diversity and complexity of users and items are fully represented. Furthermore, models without integrated high-order constraints showed a decline in metric performance compared to the complete model, indicating that adding high-order constraint information optimizes the generative process guided solely by incomplete bipartite graph data. Also, during the model training process, variants of the model experienced convergence difficulties, suggesting that AGL-SC also has significant advantages in enhancing the model's generalization capabilities.

Ultimately, AGL-SC consistently outperforms three variants, proving the effectiveness of combining variational graph reconstruction strategies with high-order feature constraints.

\subsection{Robustness Analysis}

\subsubsection{{\bfseries Data Sparsity}}

\begin{figure}
	\centering
	\subfigure[MovieLens1M]{
		\begin{minipage}[b]{0.22\textwidth}
			\includegraphics[width=1\textwidth]{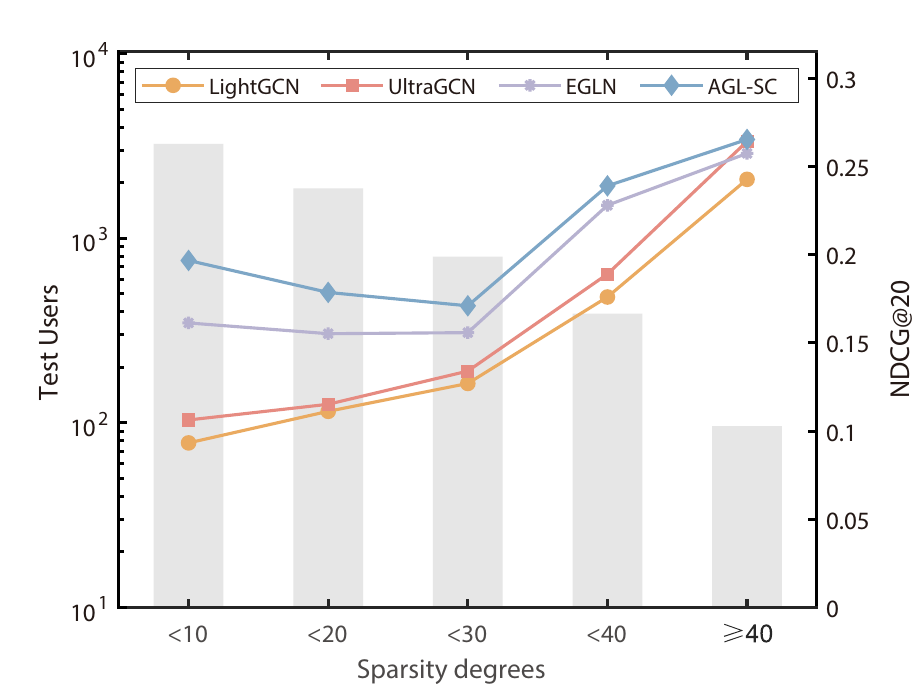}
		\end{minipage}
	}
        \subfigure[Yelp]{
    	\begin{minipage}[b]{0.22\textwidth}
   		\includegraphics[width=1\textwidth]{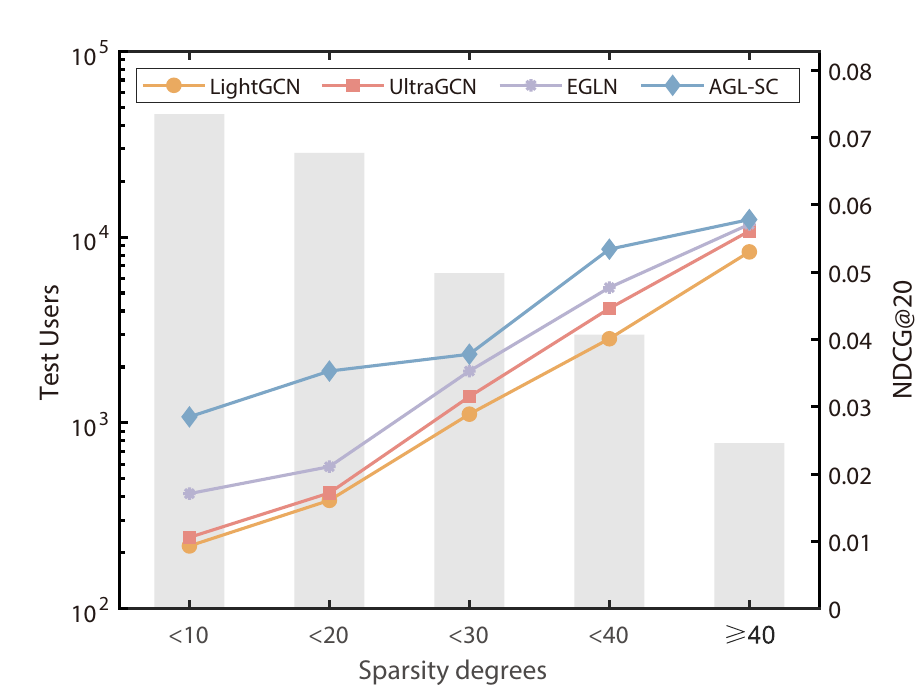}
    	\end{minipage}
    	}
	\caption{Performance comparison w.r.t different data density degrees. As the number with the x-axis increases, the sparsity of the data correspondingly diminishes.}
 \label{fig 4}

\end{figure}

\begin{figure}[ht]
  \centering
  \includegraphics[width=0.48\textwidth]{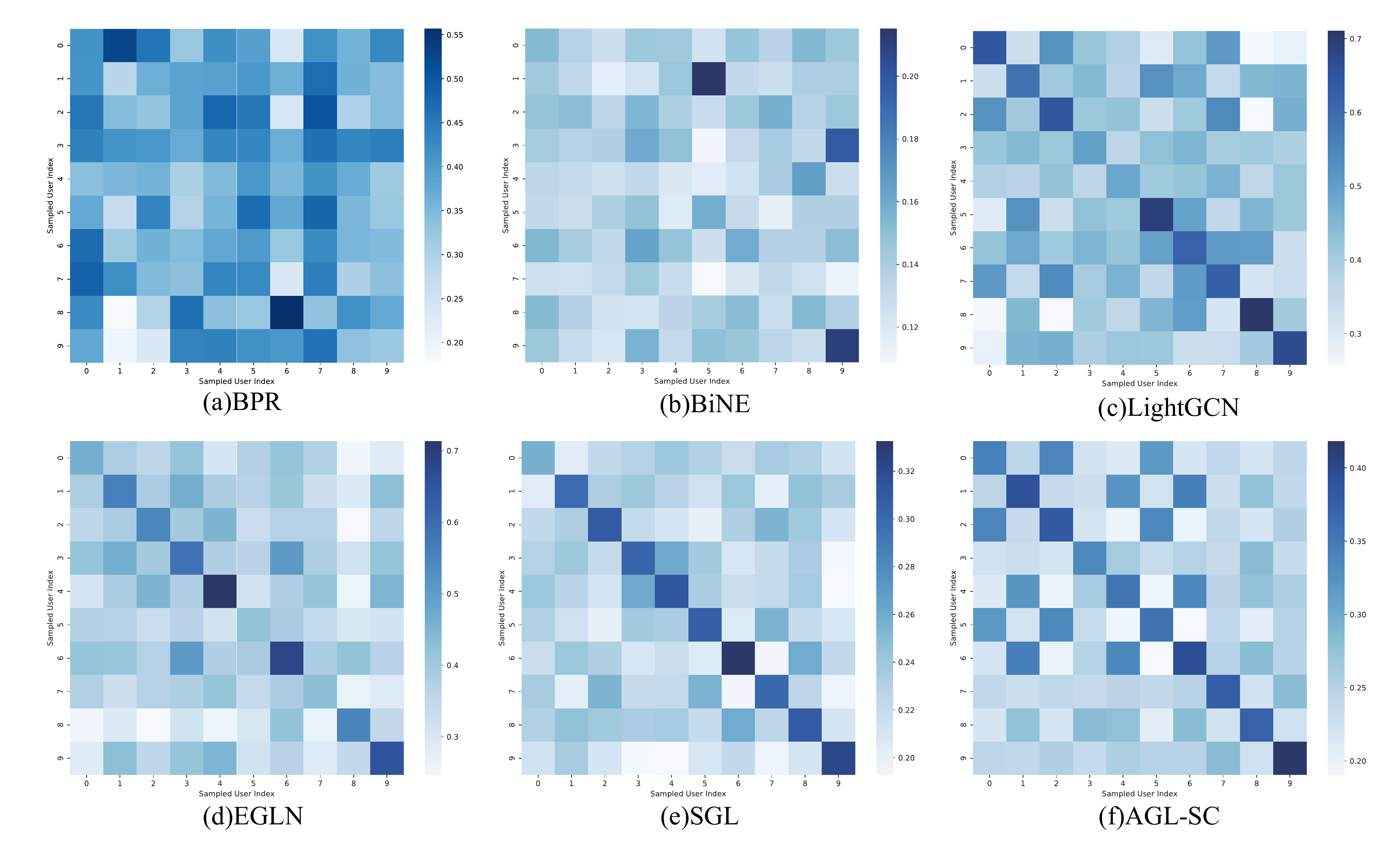}
  \caption{Comprehensive heatmap analysis of user data distribution patternsple, revealing the nuanced differences in data handling and preference recognition capabilities.}
  \label{fig 5}

\end{figure}

In recommendation systems, making recommendations to inactive users with few available interactions poses a significant challenge. Therefore, our aim is to demonstrate the effectiveness of our model in mitigating data sparsity. Specifically, we categorized test users into five groups based on sparsity, that is, the number of interactions under the target behavior (such as purchases). We then calculated the average NDCG@20 for each user group. We presented the comparative results with representative baselines on the MovieLens1M public dataset. In Fig.\ref{fig 4}, the x-axis represents different levels of data sparsity, the left side of the y-axis quantifies the number of test users in the corresponding group using bars, while the right side of the y-axis quantifies the average metric values using a line.

Based on the results, we made the following observations:
\begin{itemize}[leftmargin=*]  
  \item Compared to other methods, our model achieved better performance, especially demonstrating commendable recommendation capabilities for inactive users, who constitute a significant portion of the user base.
  \item As the number of interactions increases, the evaluation metrics slightly decline, which might be due to the increase in interaction data causing a reduction in the optimization performance of the model's objective function.
\end{itemize}

\subsubsection{{\bfseries Data Distribution}}

In this subsection, we aim to explore how the characteristics of isotropy and anisotropy affect node representations in graph learning. recommendation systems based on graph learning strive to transform nodes within a graph into low-dimensional vector representations. Such representations are instrumental in revealing complex relationships between users and items, thereby enhancing the accuracy and efficiency of recommendations. In graph learning, isotropy refers to the scenario where a node's vector representation has a similar distribution across all directions, implying that the model generates more balanced and consistent node representations to augment the capture of global information. Conversely, anisotropy denotes the presence of differences in characteristics or measurements in various directions. This allows node representations to capture a richer array of information and complex structural features.

Given the phenomenon of anisotropy in graph structures, preserving the original graph structure and its anisotropic characteristics is a primary task for the model. However, the distribution of anisotropic vector representations learned during the model's training process can lead to representation degeneration. Current algorithms counteract this by introducing isotropic noise distributions, enhancing the model's generalization capabilities, preventing overfitting, and aiding in the capture of broader contextual information. Consequently, this paper employs high-order interaction features of nodes as directed isotropic noise for perturbation, achieving a synthesis of anisotropic graph structures with enhanced isotropic structures, balancing global consistency and local differentiation in node representations. We randomly select users and items that had interactions in the original dataset and showcase their vector matrices, where a colour gradient illustrates the intensity of the interactions, with darker shades indicating a stronger preference for the items by the users. We demonstrate comparative results with representative baselines on the MovieLens1M dataset, as shown in Fig.\ref{fig 5}. The results show that, our model maintains clear boundary distinctions in terms of vector distribution similarity, while ensuring sharp contrasts between color blocks compared to other models. This indicates that our model follows an isotropic distribution, preserving structural features between different users, thereby offering sharper discrimination.

\subsection{Hyper-Parameter Sensitivities}

\begin{figure*}[htbp]
  \centering
  \includegraphics[width=1\textwidth]{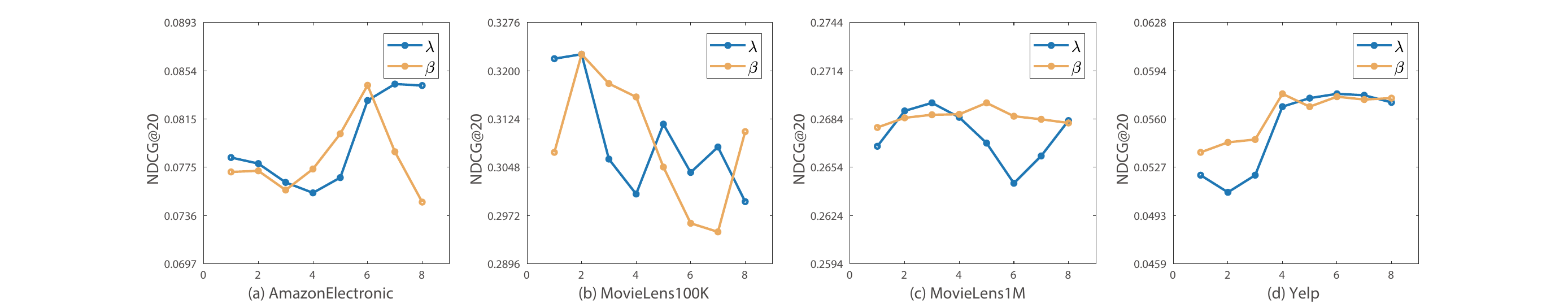}
  \caption{Performance comparison with different hyperparameter ${\lambda}$ and ${\beta}$ settings in terms of NDCG@20.}
  \label{fig 6} 
\end{figure*}

We conducted extensive experiments to examine the impact of several key hyperparameters on AGL-SC. As illustrated in Fig.\ref{fig 6}, we used the NDCG@20 metric to demonstrate the variations of three hyperparameters: the similarity constraint coefficient ${\lambda}$ and the variational inference coefficient ${\beta}$. We observed that, on the MovieLens1M dataset, AGL-SC achieved optimal performance at similarity constraint coefficient = 0.1 and variational inference coefficient = 0.1. Regarding the similarity constraint coefficient, we noticed an improvement in performance when it increased from 0 to 0.001, but a rapid decline in performance when it was set to 0.5. This indicates that appropriate similarity constraints can effectively prevent overfitting issues, but overly strong similarity constraints may limit model optimization. Similarly, for the variational inference parameter, selecting a suitable value is crucial for the overall objective optimization.

\subsection{Strategy Enhancements}

To validate the effectiveness of the proposed graph-enhanced representation learning strategy, we conducted an experiment covering two distinct datasets: MovieLens1M and Yelp. In this experiment, we utilized four classic graph neural network-based recommendation models, enhancing them with high-order constraint and generative modules to verify their performance under the NDCG@20 metric. The specific results are shown in Table \ref{table 4}.

It is evident that across both datasets, the performance of all four recommendation models improved upon the addition of the graph-enhanced representation learning strategy modules. We attribute this improvement to the fact that in recommendation systems, the latent vector of VAE is a distribution rather than fixed. This allows for a smoother differentiation in the latent space among similar samples. It also models high-order feature interactions in an implicit manner thereby enhancing the model's ability to fit user preferences. The LightGCN model showed a relatively higher degree of improvement, which we believe is due to LightGCN's simplified design, enabling it to better adapt to the integration of high-order constraint information.

\begin{table}
  \caption{Performance of different approaches under NDCG@20. "w/RA" stands for "with representation enhancement"}
  \label{table 4}
  \begin{tabular}{@{}ccccc@{}}
    \toprule
    \multirow{2}{*}{Model} & \multicolumn{2}{c}{MovieLens1M} & \multicolumn{2}{c}{Yelp}                     \\
    \cline{2-5}
    & Original & w/RA & Original & w/RA   \\
    \hline
    NGCF                   
    & 0.2457  & 0.2549(+{3.74\%}) & 0.0477   & 0.0503(+{5.45\%}) \\
    LightGCN          
    & 0.2507  & 0.2621(+{4.55\%})  & 0.0503   & 0.0541(+{7.55\%}) \\
    EGLN                  
    & 0.2574 & 0.2650(+{2.95\%}) & 0.0547   & 0.0572(+{4.57\%}) \\
    DirectAU               
    & 0.2603  & 0.2645(+{1.61\%}) & 0.0561   & 0.0575(+{2.50\%}) \\
    \bottomrule
  \end{tabular}

\end{table}

\section{CONCLUSIONS}

In this work, we introduced a graph-enhanced representation learning strategy based on sparse completion for structural enhancement and completion, aimed at reducing bias in graph node representations and addressing the issue of data sparsity in bipartite graphs within recommendation scenarios. Our method leverages the flexibility of the VAE latent space and its ability to model complex data distributions to complete missing node structures. During the generative process, high-order interaction constraints of nodes are introduced as guidance, ultimately facilitating the prediction of user preferences and item recommendations. This approach has been proven effective and superior to various advanced recommendation models across four real-world datasets. In the future, we plan to explore advanced GNN variants, such as attention-based or heterogeneous GNNs, to better adapt to the integration of multi-order constraint information.

\section{ACKNOWLEDGEMENTS}

Thanks for the support provided by MindSpore Community.

\newpage
\bibliographystyle{IEEEtran}
\bibliography{IEEEabrv,sample-sigconf}

\begin{thebibliography}{10}
\providecommand{\url}[1]{#1}
\csname url@samestyle\endcsname
\providecommand{\newblock}{\relax}
\providecommand{\bibinfo}[2]{#2}
\providecommand{\BIBentrySTDinterwordspacing}{\spaceskip=0pt\relax}
\providecommand{\BIBentryALTinterwordstretchfactor}{4}
\providecommand{\BIBentryALTinterwordspacing}{\spaceskip=\fontdimen2\font plus
\BIBentryALTinterwordstretchfactor\fontdimen3\font minus \fontdimen4\font\relax}
\providecommand{\BIBforeignlanguage}[2]{{%
\expandafter\ifx\csname l@#1\endcsname\relax
\typeout{** WARNING: IEEEtran.bst: No hyphenation pattern has been}%
\typeout{** loaded for the language `#1'. Using the pattern for}%
\typeout{** the default language instead.}%
\else
\language=\csname l@#1\endcsname
\fi
#2}}
\providecommand{\BIBdecl}{\relax}
\BIBdecl

\bibitem{wu2022survey}
L.~Wu, X.~He, X.~Wang, K.~Zhang, and M.~Wang, ``A survey on accuracy-oriented neural recommendation: From collaborative filtering to information-rich recommendation,'' \emph{IEEE Transactions on Knowledge and Data Engineering}, vol.~35, no.~5, pp. 4425--4445, 2022.

\bibitem{ye2023grace}
H.~Ye, S.~Vedula, Y.~Chen, Y.~Yang, A.~Bronstein, R.~Dreslinski, T.~Mudge, and N.~Talati, ``Grace: A scalable graph-based approach to accelerating recommendation model inference,'' in \emph{Proceedings of the 28th ACM International Conference on Architectural Support for Programming Languages and Operating Systems, Volume 3}, 2023, pp. 282--301.

\bibitem{xie2021deep}
R.~Xie, C.~Ling, Y.~Wang, R.~Wang, F.~Xia, and L.~Lin, ``Deep feedback network for recommendation,'' in \emph{Proceedings of the Twenty-Ninth International Conference on International Joint Conferences on Artificial Intelligence}, 2021, pp. 2519--2525.

\bibitem{chen2022review}
Z.~Chen and S.~Wang, ``A review on matrix completion for recommender systems,'' \emph{Knowledge and Information Systems}, pp. 1--34, 2022.

\bibitem{farias2022uncertainty}
V.~Farias, A.~A. Li, and T.~Peng, ``Uncertainty quantification for low-rank matrix completion with heterogeneous and sub-exponential noise,'' in \emph{International Conference on Artificial Intelligence and Statistics}.\hskip 1em plus 0.5em minus 0.4em\relax PMLR, 2022, pp. 1179--1189.

\bibitem{he2024kgcna}
G.~He, Z.~Zhang, H.~Wu, S.~Luo, and Y.~Liu, ``Kgcna: Knowledge graph collaborative neighbor awareness network for recommendation,'' \emph{IEEE Transactions on Emerging Topics in Computational Intelligence}, 2024.

\bibitem{Zhang2020Inductive}
M.~Zhang and Y.~Chen, ``Inductive matrix completion based on graph neural networks,'' in \emph{International Conference on Learning Representations}, 2020.

\bibitem{yang2021enhanced}
Y.~Yang, L.~Wu, R.~Hong, K.~Zhang, and M.~Wang, ``Enhanced graph learning for collaborative filtering via mutual information maximization,'' in \emph{Proceedings of the 44th International ACM SIGIR Conference on Research and Development in Information Retrieval}, 2021, pp. 71--80.

\bibitem{dai2020usual}
B.~Dai, Z.~Wang, and D.~Wipf, ``The usual suspects? reassessing blame for vae posterior collapse,'' in \emph{International conference on machine learning}.\hskip 1em plus 0.5em minus 0.4em\relax PMLR, 2020, pp. 2313--2322.

\bibitem{he2022masked}
K.~He, X.~Chen, S.~Xie, Y.~Li, P.~Doll{\'a}r, and R.~Girshick, ``Masked autoencoders are scalable vision learners,'' in \emph{Proceedings of the IEEE/CVF conference on computer vision and pattern recognition}, 2022, pp. 16\,000--16\,009.

\bibitem{wu2020diffnet++}
L.~Wu, J.~Li, P.~Sun, R.~Hong, Y.~Ge, and M.~Wang, ``Diffnet++: A neural influence and interest diffusion network for social recommendation,'' \emph{IEEE Transactions on Knowledge and Data Engineering}, vol.~34, no.~10, pp. 4753--4766, 2020.

\bibitem{nakagawa2022gromov}
N.~Nakagawa, R.~Togo, T.~Ogawa, and M.~Haseyama, ``Gromov-wasserstein autoencoders,'' in \emph{The Eleventh International Conference on Learning Representations}, 2022.

\bibitem{li2023generalized}
Y.~Li, X.~Wang, H.~Liu, and C.~Shi, ``A generalized neural diffusion framework on graphs,'' in \emph{Proceedings of the AAAI Conference on Artificial Intelligence}, vol.~38, no.~8, 2024, pp. 8707--8715.

\bibitem{he2017neural2}
X.~He and T.-S. Chua, ``Neural factorization machines for sparse predictive analytics,'' in \emph{Proceedings of the 40th International ACM SIGIR conference on Research and Development in Information Retrieval}, 2017, pp. 355--364.

\bibitem{perozzi2014deepwalk}
B.~Perozzi, R.~Al-Rfou, and S.~Skiena, ``Deepwalk: Online learning of social representations,'' in \emph{Proceedings of the 20th ACM SIGKDD international conference on Knowledge discovery and data mining}, 2014, pp. 701--710.

\bibitem{grover2016node2vec}
A.~Grover and J.~Leskovec, ``node2vec: Scalable feature learning for networks,'' in \emph{Proceedings of the 22nd ACM SIGKDD international conference on Knowledge discovery and data mining}, 2016, pp. 855--864.

\bibitem{shi2021sgcn}
L.~Shi, L.~Wang, C.~Long, S.~Zhou, M.~Zhou, Z.~Niu, and G.~Hua, ``Sgcn: Sparse graph convolution network for pedestrian trajectory prediction,'' in \emph{Proceedings of the IEEE/CVF Conference on Computer Vision and Pattern Recognition}, 2021, pp. 8994--9003.

\bibitem{ren2023disentangled}
X.~Ren, L.~Xia, J.~Zhao, D.~Yin, and C.~Huang, ``Disentangled contrastive collaborative filtering,'' in \emph{Proceedings of the 46th International ACM SIGIR Conference on Research and Development in Information Retrieval}, 2023, pp. 1137--1146.

\bibitem{yang2023generative}
Y.~Yang, Z.~Wu, L.~Wu, K.~Zhang, R.~Hong, Z.~Zhang, J.~Zhou, and M.~Wang, ``Generative-contrastive graph learning for recommendation,'' in \emph{Proceedings of the 46th International ACM SIGIR Conference on Research and Development in Information Retrieval}, 2023, pp. 1117--1126.

\bibitem{ying2018graph}
R.~Ying, R.~He, K.~Chen, P.~Eksombatchai, W.~L. Hamilton, and J.~Leskovec, ``Graph convolutional neural networks for web-scale recommender systems,'' in \emph{Proceedings of the 24th ACM SIGKDD international conference on knowledge discovery \& data mining}, 2018, pp. 974--983.

\bibitem{wang2019neural}
X.~Wang, X.~He, M.~Wang, F.~Feng, and T.-S. Chua, ``Neural graph collaborative filtering,'' in \emph{Proceedings of the 42nd international ACM SIGIR conference on Research and development in Information Retrieval}, 2019, pp. 165--174.

\bibitem{he2020lightgcn}
X.~He, K.~Deng, X.~Wang, Y.~Li, Y.~Zhang, and M.~Wang, ``Lightgcn: Simplifying and powering graph convolution network for recommendation,'' in \emph{Proceedings of the 43rd International ACM SIGIR conference on research and development in Information Retrieval}, 2020, pp. 639--648.

\bibitem{mao2021ultragcn}
K.~Mao, J.~Zhu, X.~Xiao, B.~Lu, Z.~Wang, and X.~He, ``Ultragcn: ultra simplification of graph convolutional networks for recommendation,'' in \emph{Proceedings of the 30th ACM International Conference on Information \& Knowledge Management}, 2021, pp. 1253--1262.

\bibitem{peng2022svd}
S.~Peng, K.~Sugiyama, and T.~Mine, ``Svd-gcn: A simplified graph convolution paradigm for recommendation,'' in \emph{Proceedings of the 31st ACM International Conference on Information \& Knowledge Management}, 2022, pp. 1625--1634.

\bibitem{chen2020efficient}
C.~Chen, M.~Zhang, Y.~Zhang, Y.~Liu, and S.~Ma, ``Efficient neural matrix factorization without sampling for recommendation,'' \emph{ACM Transactions on Information Systems (TOIS)}, vol.~38, no.~2, pp. 1--28, 2020.

\bibitem{cao2021bipartite}
J.~Cao, X.~Lin, S.~Guo, L.~Liu, T.~Liu, and B.~Wang, ``Bipartite graph embedding via mutual information maximization,'' in \emph{Proceedings of the 14th ACM international conference on web search and data mining}, 2021, pp. 635--643.

\bibitem{monti2017geometric}
F.~Monti, M.~Bronstein, and X.~Bresson, ``Geometric matrix completion with recurrent multi-graph neural networks,'' \emph{Advances in neural information processing systems}, vol.~30, 2017.

\bibitem{berg2017graph}
R.~v.~d. Berg, T.~N. Kipf, and M.~Welling, ``Graph convolutional matrix completion,'' \emph{arXiv preprint arXiv:1706.02263}, 2017.

\bibitem{nguyen2019geometric}
D.~M. Nguyen, R.~Calderbank, and N.~Deligiannis, ``Geometric matrix completion with deep conditional random fields,'' \emph{IEEE Transactions on Neural Networks and Learning Systems}, vol.~31, no.~9, pp. 3579--3593, 2019.

\bibitem{wu2020joint}
L.~Wu, Y.~Yang, K.~Zhang, R.~Hong, Y.~Fu, and M.~Wang, ``Joint item recommendation and attribute inference: An adaptive graph convolutional network approach,'' in \emph{Proceedings of the 43rd International ACM SIGIR conference on research and development in Information Retrieval}, 2020, pp. 679--688.

\bibitem{zhang2022mc2g}
Q.~Zhang, G.~Suh, C.~Suh, and V.~Y. Tan, ``Mc2g: An efficient algorithm for matrix completion with social and item similarity graphs,'' \emph{IEEE Transactions on Signal Processing}, vol.~70, pp. 2681--2697, 2022.

\bibitem{wang2023diffusion}
W.~Wang, Y.~Xu, F.~Feng, X.~Lin, X.~He, and T.-S. Chua, ``Diffusion recommender model,'' in \emph{Proceedings of the 46th International ACM SIGIR Conference on Research and Development in Information Retrieval}, 2023, pp. 832--841.

\bibitem{yang2023generate}
Z.~Yang, J.~Wu, Z.~Wang, X.~Wang, Y.~Yuan, and X.~He, ``Generate what you prefer: Reshaping sequential recommendation via guided diffusion,'' \emph{Advances in Neural Information Processing Systems}, vol.~36, 2024.

\bibitem{ye2023graph}
Y.~Ye, L.~Xia, and C.~Huang, ``Graph masked autoencoder for sequential recommendation,'' in \emph{Proceedings of the 46th International ACM SIGIR Conference on Research and Development in Information Retrieval}, 2023, pp. 321--330.

\bibitem{liang2018variational}
D.~Liang, R.~G. Krishnan, M.~D. Hoffman, and T.~Jebara, ``Variational autoencoders for collaborative filtering,'' in \emph{Proceedings of the 2018 world wide web conference}, 2018, pp. 689--698.

\bibitem{wang2023collaboration}
Y.~Wang, Y.~Zhao, Y.~Zhang, and T.~Derr, ``Collaboration-aware graph convolutional network for recommender systems,'' in \emph{Proceedings of the ACM Web Conference 2023}, 2023, pp. 91--101.

\bibitem{jiang2019semi}
B.~Jiang, Z.~Zhang, D.~Lin, J.~Tang, and B.~Luo, ``Semi-supervised learning with graph learning-convolutional networks,'' in \emph{Proceedings of the IEEE/CVF conference on computer vision and pattern recognition}, 2019, pp. 11\,313--11\,320.

\bibitem{ma2019learning}
J.~Ma, C.~Zhou, P.~Cui, H.~Yang, and W.~Zhu, ``Learning disentangled representations for recommendation,'' \emph{Advances in neural information processing systems}, vol.~32, 2019.

\bibitem{shenbin2020recvae}
I.~Shenbin, A.~Alekseev, E.~Tutubalina, V.~Malykh, and S.~I. Nikolenko, ``Recvae: A new variational autoencoder for top-n recommendations with implicit feedback,'' in \emph{Proceedings of the 13th international conference on web search and data mining}, 2020, pp. 528--536.

\bibitem{tolstikhin2017wasserstein}
I.~Tolstikhin, O.~Bousquet, S.~Gelly, and B.~Sch{\"o}lkopf, ``Wasserstein auto-encoders,'' in \emph{6th International Conference on Learning Representations (ICLR 2018)}, 2018.

\end{thebibliography}
\end{document}